\documentclass[aps,prl,reprint,preprintnumbers,superscriptaddress,amsmath,amssymb,bibnotes,longbibliography]{revtex4-2}

\usepackage{graphicx}% Include figure files
\usepackage{bm}% bold math
\usepackage[colorlinks,linkcolor=blue,anchorcolor=blue,citecolor=blue,urlcolor=blue,filecolor=blue,menucolor=blue,runcolor=blue]{hyperref}% add hypertext capabilities
\graphicspath{{figures/}}

\begin{document}
	
	%\preprint{Draft Bin Shen}

	\title{Magnetic field and pressure tuning of the heavy fermion antiferromagnet CePdIn}% Force line breaks with \\
	
	\author{Bin Shen}
	\email{bin.shen@physik.uni-augsburg.de}
	\altaffiliation{Present address: Experimental Physics VI, Center for Electronic Correlations and Magnetism, University of Augsburg, 86159 Augsburg, Germany}
	\affiliation  {New Cornerstone Science Laboratory, Center for Correlated Matter and School of Physics, Zhejiang University, Hangzhou 310058, China}
	
	\author{Feng Du}
	\altaffiliation{Max Planck Institute for Chemistry, Hahn Meitner Weg 1, Mainz 55128, Germany}
	\affiliation  {New Cornerstone Science Laboratory, Center for Correlated Matter and School of Physics, Zhejiang University, Hangzhou 310058, China}
	
	\author{Rui Li}
	\affiliation  {New Cornerstone Science Laboratory, Center for Correlated Matter and School of Physics, Zhejiang University, Hangzhou 310058, China}
	
	\author{Hang Su}
	\altaffiliation{Department of Physics, University of Tokyo, Bunkyo-ku, Tokyo, Japan}
	\affiliation  {New Cornerstone Science Laboratory, Center for Correlated Matter and School of Physics, Zhejiang University, Hangzhou 310058, China}
	
	\author{Yasuyuki Shimura}
	\affiliation  {Department of Quantum Matter, Graduate School of Advanced Science and Engineering, Hiroshima University, Higashi-Hiroshima 739-8530, Japan}
	
	\author{Takahiro Onimaru}
	\affiliation  {Department of Quantum Matter, Graduate School of Advanced Science and Engineering, Hiroshima University, Higashi-Hiroshima 739-8530, Japan}

	\author{Kazunori Umeo}
	\affiliation  {Department of Low Temperature Experiment, Integral Experimental Support/Research Division, N-BARD, Hiroshima University, Higashi-Hiroshima 739-8526, Japan}
	
	\author{Xin Lu}
	\affiliation {New Cornerstone Science Laboratory, Center for Correlated Matter and School of Physics, Zhejiang University, Hangzhou 310058, China}
	
	\author{Toshiro Takabatake}
	\affiliation  {Department of Quantum Matter, Graduate School of Advanced Science and Engineering, Hiroshima University, Higashi-Hiroshima 739-8530, Japan}
	\affiliation  {New Cornerstone Science Laboratory, Center for Correlated Matter and School of Physics, Zhejiang University, Hangzhou 310058, China}
	
	\author{Michael Smidman}
	\email{msmidman@zju.edu.cn}
	\affiliation  {New Cornerstone Science Laboratory, Center for Correlated Matter and School of Physics, Zhejiang University, Hangzhou 310058, China}

	\author{Huiqiu Yuan}
	\email{hqyuan@zju.edu.cn}
	\affiliation{New Cornerstone Science Laboratory, Center for Correlated Matter and School of Physics, Zhejiang University, Hangzhou 310058, China}
	\affiliation{Institute for Advanced Study in Physics, Zhejiang University, Hangzhou 310058, China}
	\affiliation{Institute of Fundamental and Transdisciplinary Research, Zhejiang University, Hangzhou 310058, China}
	\affiliation{State Key Laboratory of Silicon and Advanced Semiconductor Materials, Zhejiang University, Hangzhou 310058, China}


	\begin{abstract}
		Frustrated Kondo lattices are ideal platforms for studying how both the Kondo effect and quantum fluctuations compete with the magnetic exchange interactions that drive magnetic ordering. Here, we investigate the effect of tuning the heavy-fermion compound CePdIn, which crystallizes in the geometrically frustrated ZrNiAl-type structure, using applied magnetic fields and hydrostatic pressure. At ambient pressure, CePdIn exhibits two magnetic transitions, one at $T_{\rm{N}} \approx 1.65$~K and another at $T_{\rm{M}} \approx 1.15$~K, which are both suppressed by applied $c$-axis fields. Upon applying pressure in zero magnetic field, there is a non-monotonic evolution of $T_{\rm{N}}$, which decreases  to 0.8~K at 2.3~GPa, before abruptly increasing to ~1.5~K at 2.6~GPa. At higher pressures, $T_{\rm{N}}$ has a weak pressure dependence, and vanishes near 5 GPa. Together with the high-pressure phase being more robust to applied fields, these results suggest two distinct antiferromagnetic phases in CePdIn, which are separated near 2.6 GPa, and this change may be driven by the evolution of the underlying electronic structure due to enhanced Kondo hybridization under pressure.
	\end{abstract}
	
	%pacs{Valid PACS appear here}% PACS, the Physics and Astronomy
	\maketitle
	
	%\tableofcontents
	
	\section{I. Introduction}
	
	Quantum fluctuations driven by magnetic frustration or reduced dimensionality can stabilize correlated zero-temperature states such as spin liquids in insulating materials \cite{SL2010}. In contrast, heavy fermion compounds host mobile electrons, introducing two competing mechanisms that govern their magnetic properties: the Kondo effect and the Ruderman–Kittel–Kasuya–Yosida (RKKY) interaction. The Kondo effect leads to an onsite screening of local magnetic moments by the conduction electrons, favoring a non-magnetic ground state, while the RKKY interaction favors magnetic order due to the conduction electrons mediating magnetic exchange. This competition makes heavy fermion systems highly tunable, with their ground states sensitive to external parameters \cite{Gegenwart2008}. Owing to their low energy scales and tunable ground states, heavy fermion compounds serve as an ideal platform for probing the influence of quantum fluctuations on their physical properties. It is of particular interest to examine their properties when both magnetic frustration and the Kondo effect have a significant influence \cite{PBqs2006, JLTPColeman2010,Chen2024}.     
	
	Heavy fermion materials with the hexagonal ZrNiAl-type structure are promising systems for examining the interplay of magnetic frustration and the Kondo effect, where the rare-earth moments form a distorted kagome lattice in the $ab$-plane ~\cite{JJAPShf1987,JPBkh1994,Fujita1998,Xie2022}. This is exemplified by CePdAl which exhibits an antiferromagnetic transition at $T_{\rm{M}} = 2.7$~K~\cite{JPBkh1994}, and the presence of strong frustration is revealed by neutron diffraction measurements showing that one-third of the Ce moments remain disordered down to at least 30~mK~\cite{JPCMad1996}. Novel forms of quantum criticality have also been revealed in CePdAl upon applying magnetic fields \cite{PRLsl2017,PRBzh2016}, doping \cite{Sakai2016,Ishant2024} and pressure \cite{NP2019CPA,Majumder2022}. Pressurization induces an extended quantum critical phase together with strange metal behavior, likely driven by the strong frustration \cite{NP2019CPA,Majumder2022}. Among isostructural compounds, YbPtIn also exhibits strange-metal like behaviors at a field induced quantum critical point  \cite{PRLme2006}, while a low-temperature bicritical point is revealed in YbAgGe \cite{PRLty2013}, that is associated with amplitude modulated magnetically ordered phases \cite{Larsen2021}. Meanwhile, CeRhSn does not order at low temperatures, and instead exhibits ambient pressure quantum criticality in the presence of significant frustration \cite{SAyt2015,Kittaka2021,Tripathi2022,Kimura2025}.
	
	We focus on CePdIn, where In is an isovalent element with Al. CePdIn exhibits an antiferromagnetic phase transition at $T_{\rm{N}} \approx$ 1.7~K~\cite{AIPeb1988,JAChf1992,SSClg2007}, followed by  an additional phase transition at $T_{\rm{M}} \approx$ 1.0~K~\cite{JACbe1993}. The zero-temperature magnetic specific heat $C_{\rm{mag}}/T$ reaches a substantial value of 1.4~J mol$^{-1}$ K$^{-2}$~\cite{Maeno_1987}. Chemical pressure studies via Ni-doping in CeNi$_x$Pd$_{1-x}$In reveal a systematic suppression of $T_{\rm{N}}$ with increasing $x$, whereby non-Fermi liquid behavior is observed in the composition range $0.4 \leq x \leq 0.6$~\cite{JPCMmk2013}. Under moderate hydrostatic pressures up to 1.9 GPa, resistivity measurements down to 4.2 K showed a monotonic increase in the magnetic resistivity  \cite{Kurisu1990}.  Compared to CePdAl, which has lattice parameters $a = 7.220$~\AA~and $c = 4.233$~\AA~\cite{JAChf1993}, CePdIn with $a = 7.698$~\AA, and $c = 4.076$~\AA~exhibits an expanded $a$-axis and compressed $c$-axis~\cite{AIPeb1988}, which may enhance the interplane magnetic interactions relative to those in-plane, potentially resulting in a more three-dimensional magnetism with weaker geometric frustration. Given that it is of particular interest to determine the key parameters governing frustration together with the resulting quantum critical behaviors in heavy fermion magnets, here we examine the effects of tuning CePdIn with applied magnetic fields and pressure.

	\section{II. experimental methods}
	
	Single-crystal CePdIn samples were synthesized using the Czochralski method. The constituent elements were melted in stoichiometric proportions using a radio-frequency induction furnace, and this boule was used as the starting material for pulling a crystal using a tetra-arc furnace, yielding properties similar to those reported in Ref.~\cite{PBhf1990}. The temperature dependence of the magnetic susceptibility above 2~K and magnetization as a function of field were measured in a Quantum Design Magnetic Property Measurement System (MPMS) with a helium-3 insert (0.4~K and 1.2~K). Magnetization as a function of field at 0.1~K was measured by the Faraday method with a homemade capacitive magnetometer~\cite{21ShiRSI}, mounted on a dilution refrigerator. The temperature dependence of the specific heat down to 0.4~K was measured in a Quantum Design Physical Property Measurement System (PPMS) with a $^3$He insert, using a standard  relaxation method.  The resistivity both at ambient pressure and under pressures in  piston-cylinder pressure cells was measured using the standard four probe method. The measurements from 4~K down to 0.3~K were performed in an Oxford Instruments $^3$He cryostat with a 9~T magnet, and the high temperature parts of the resistivity above 1.9~K were measured in a Quantum Design PPMS. Ac heat capacity under pressure was measured via an ac calorimetric technique \cite{acC}. Pressures in the piston-cylinder cell were determined by measuring the superconducting transition temperature of Pb~\cite{81EilJPM}. Daphne 7373 was used as the pressure-transmitting medium.

	Measurements of the resistivity and heat capacity in a diamond anvil cell (DAC) were performed in a Teslatron-PT system with an Oxford $^3$He refrigerator, in a temperature range of 0.3~K to 300~K, in applied magnetic fields up to 8~T.  Single crystals of CePdIn were polished and cut into rectangular pieces with approximate dimensions of 200~$\mu$m $\times$ 80~$\mu$m $\times$ 40~$\mu$m, loaded into a BeCu DAC with a 1000-$\mu$m-diameter culet. A  100-$\mu$m-thick preindented CuBe gasket was covered with Al$_2$O$_3$ for electrical insulation and a 500-$\mu$m-diameter hole was drilled as the sample chamber. Daphne 7373 was used as the pressure-transmitting medium. Ruby chips were loaded into the DAC for determining the pressure by the ruby fluorescence method at room temperature~\cite{86MaoJGR}. For electrical transport measurements, four 15~$\mu$m-diameter gold wires were glued to the samples with silver epoxy paste and the resistivity was measured using the standard four-probe method. Adapted ac-calorimetry measurements were also performed in the DAC using Au(Fe)-Au as a thermocouple and via self-heating by injecting an ac current to the sample as an oscillating heating power source \cite{02WilJPCM}.

	\section{III. results}
	
	\subsection{A. Magnetic and transport properties at ambient pressure}
	
	\begin{figure}
		\includegraphics[angle=0,width=0.45\textwidth]{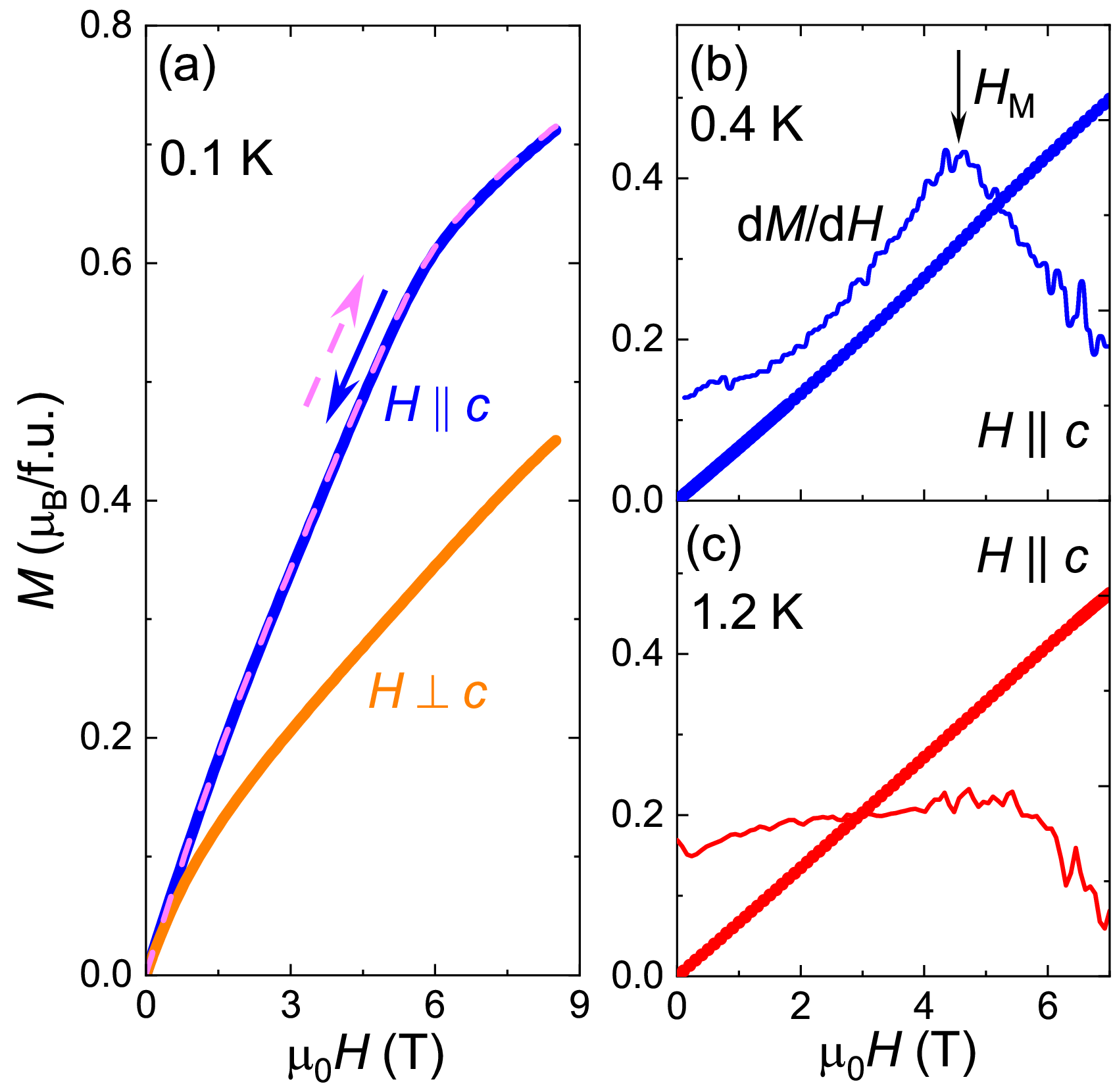}
		\vspace{-12pt} \caption{\label{MHall} Isothermal magnetization versus applied field $M(H)$ of CePdIn at (a) 0.1 K for magnetic fields along and perpendicular to the $c$-axis,  as well as at (b) 0.4~K, and (c) 1.2~K for magnetic fields along the $c$-axis. The solid lines in (b) and (c) are the field derivatives of $M(H)$. Arrows in (a) indicate whether the measurement was performed upon increasing or decreasing the field. Arrow in (b) indicates the metamagnetic transition $H_{\rm{M}}$.}
		\vspace{-12pt}
	\end{figure}
	
	The isothermal magnetization versus field $M(H)$ of CePdIn below $T_{\rm{N}}$ is shown in Fig.~\ref{MHall}. At 0.1~K [Fig.~\ref{MHall} (a)], the magnetization reveals isotropic behaviorbelow 1~T. With increasing magnetic field, the anisotropy becomes more pronounced, reaching $M_{\rm{c}}/M_{\rm{ab}} \approx 1.58$ at 8.5 T. Reversing the field direction reveals no detectable hysteresis. A kink feature appears at approximately 6~T for fields applied along the $c$-axis, which may correspond to the transition from an antiferromagnetic to paramagnetic state. At an elevated temperature of 1.2~K [Fig.~\ref{MHall} (c)], $M(H)$ exhibits a nearly linear field dependence up to 7~T. While the $M(H)$ at 0.4~K exhibits a similar increase as shown in Fig.~\ref{MHall} (b), the derivative $dM/dH$ exhibits a clear maximum, indicating a weak field-induced metamagnetic transition at 4.5~T, likely associated with the lower transition $T_{\rm{M}}$. This behavior contrasts sharply with isostructural CePdAl, where magnetization measurements show multiple metamagnetic transitions in the magnetically ordered phase \cite{02GotJPCS, PRLsl2017,PRBzh2016}.

	\begin{figure}
		\includegraphics[angle=0,width=0.45\textwidth]{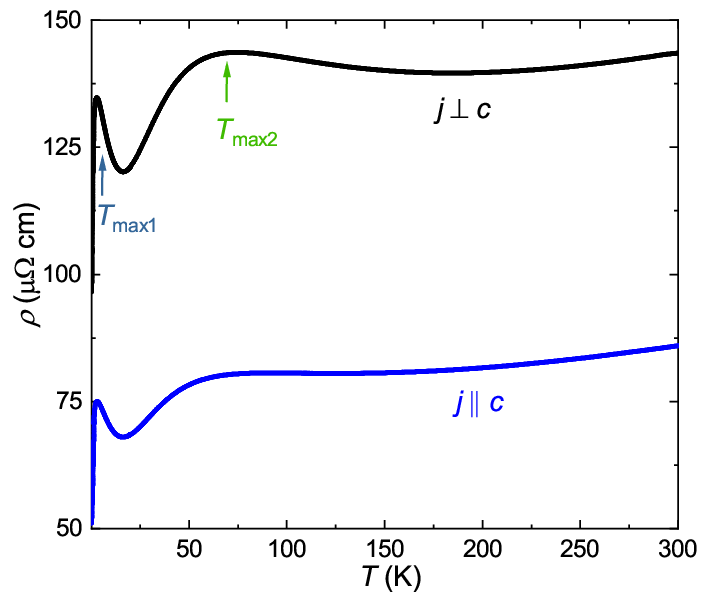}
		\vspace{-12pt} \caption{\label{RT} Temperature dependence of the resistivity $\rho(T)$   of CePdIn with the current applied parallel and perpendicular to the $c$-axis. The arrows indicates the temperature corresponding to the resistivity maxima.}
		\vspace{-12pt}
	\end{figure} 
	
	The temperature dependence of the  resistivity, shown in Fig.~\ref{RT}, exhibits characteristic behaviors of a Ce-based Kondo lattice. Both the measurements with $c$-axis ($\rho_c$) and $ab$-plane ($\rho_{ab}$) currents exhibit two maxima, which are typical features arising from the interplay between Kondo scattering and crystalline electric field effects. The low-temperature maximum occurs at $T_{\rm{max1}} \approx 3$~K, in good agreement with the Kondo temperature $T_{\rm{K}} = 3.3$~K estimated from the analysis of the entropy~\cite{JPCks1988}. A second, broader maximum appears at $T_{\rm{max2}} \approx 70$~K. The resistivity anisotropy $\rho_{ab}/\rho_c \approx 2$ is consistent with previous single-crystal studies~\cite{PBhf1990}, confirming the higher conductivity along the $c$-axis.

	\begin{figure}
		\includegraphics[angle=0,width=0.45\textwidth]{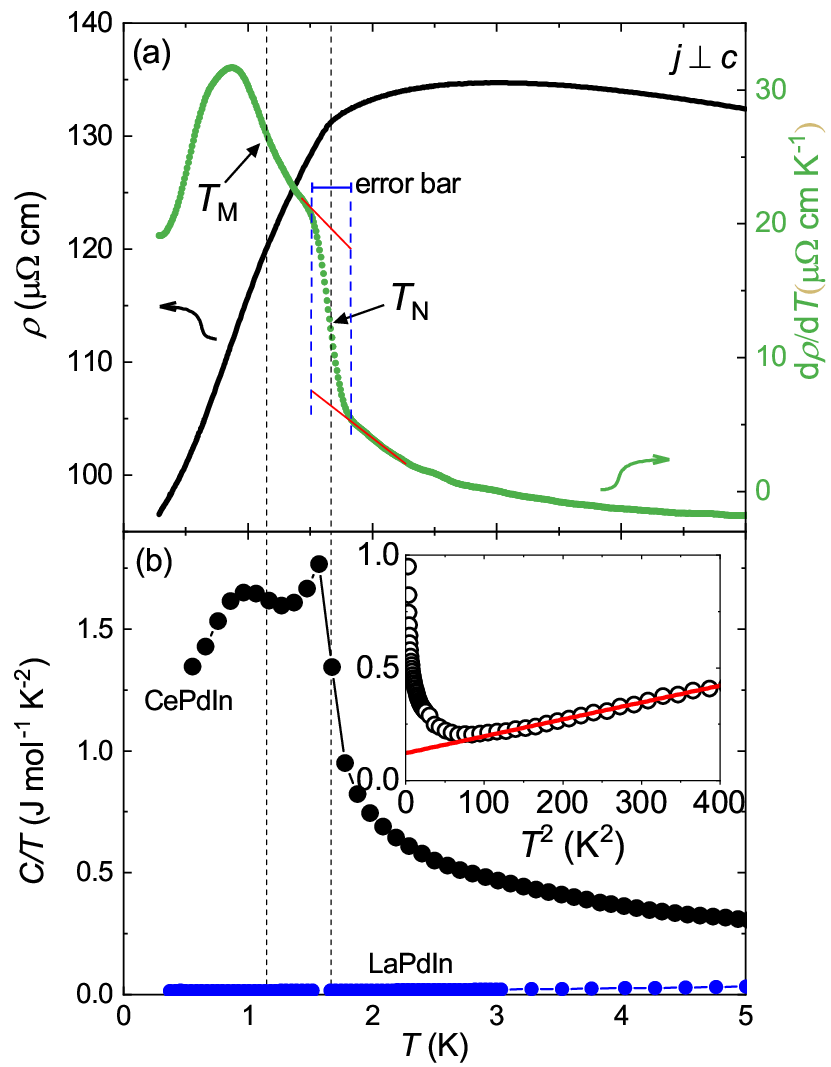}
		\vspace{-12pt} \caption{\label{RCT} (a) Temperature dependence of the resistivity $\rho(T)$  and  its temperature derivative d$\rho(T)$/d$T$ with the current applied perpendicular to the $c$-axis. The arrows mark the two transitions at $T_{\rm{N}}$ and $T_{\rm{M}}$. The error bar reflects the finite width of the transition. (b) Temperature dependence of the specific heat as $C(T)/T$ of CePdIn and LaPdIn. The inset displays a plot of  $C/T$ versus $T^2$, and the solid line represents a linear fit. }
		\vspace{-12pt}
	\end{figure}  
	
	At low temperatures, two distinct anomalies emerge in both the resistivity and heat capacity at $T_{\rm{N}}$ = 1.65~K and $T_{\rm{M}}$ = 1.15 K (Fig.~\ref{RCT}), corresponding to the previously reported phase transitions~\cite{JACbe1993}. Above $T_{\rm{N}}$, $C/T$ exhibits a pronounced increase upon cooling, signaling the development of heavy quasiparticles or strong spin fluctuations. At higher temperatures, the specific heat of CePdIn follows  conventional metallic behavior $C/T = \gamma + \beta T^2$, yielding $\gamma = 121$~mJ/mol\,K$^2$, which corresponds to a moderately enhanced value, as compared to the non-magnetic analog LaPdIn which has $\gamma = 11$~mJ/mol\,K$^2$. 
	
	\begin{figure}
		\includegraphics[angle=0,width=0.49\textwidth]{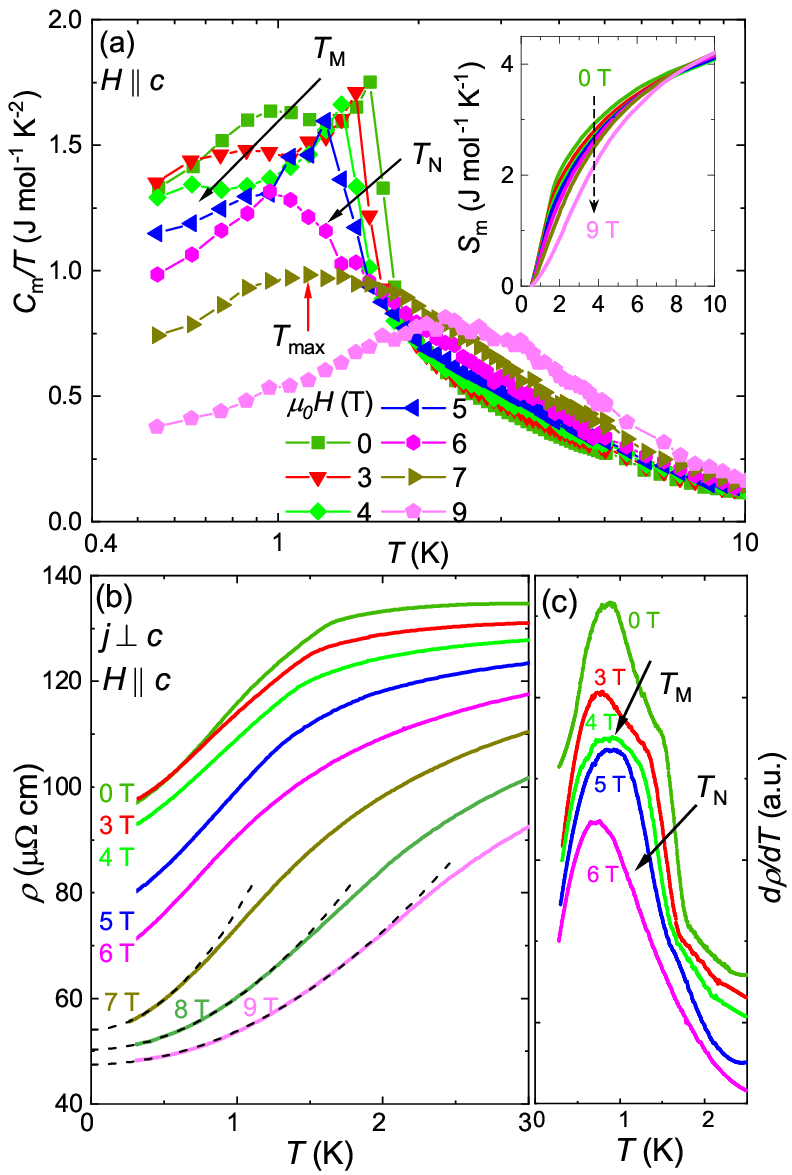}
		\vspace{-12pt} \caption{\label{CRFiled} (a) Temperature dependence of the magnetic contribution to the specific heat as $C_{\rm{m}}(T)/T$ on a logarithmic temperature scale, with various magnetic fields applied along the $c$-axis. The inset shows the temperature dependence of the magnetic entropy $S_{\rm{m}}(T)$ at various magnetic fields. (b) Temperature dependence of the resistivity $\rho(T)$ in various applied $c$-axis magnetic fields.  The dashed lines correspond to fitting with Fermi liquid behavior. (c) Temperature derivative of the resistivity d$\rho(T)$/d$T$ in various applied $c$-axis  magnetic fields. The black arrows in (a) and (c) highlight the evolution of the transition temperatures $T_{\rm{N}}$ and $T_{\rm{M}}$ with field, and the red arrow in (a) marks the hump temperature $T_{\rm{max}}$ in the heat capacity.}
		\vspace{-12pt}
	\end{figure} 
	
	Figure~\ref{CRFiled}(a) displays the magnetic contribution to the specific heat, $C_{\rm{m}}/T$ with different magnetic fields applied along the $c$-axis, obtained by subtracting the LaPdIn data from that of CePdIn. With increasing field, both $T_{\rm{N}}$ and $T_{\rm{M}}$  shift to lower temperatures, and for fields above 6~T, the sharp transitions in $C_{\rm{m}}/T$ are replaced by a broad hump that moves to higher temperatures with increasing field, indicating the disappearance of antiferromagnetic order. The inset of Fig.~\ref{CRFiled}(a) shows the temperature dependence of the magnetic entropy under various fields. At 10~K, the entropy reaches approximately 70\% of $R$ ln 2. Upon entering the magnetically ordered phase, the entropy is reduced to about 35\% of $R$ ln 2, indicating a substantial degree of Kondo hybridization. Under applied magnetic fields, the low-temperature entropy decreases monotonically, showing no enhancement of magnetic fluctuations and thus providing no evidence for a field-induced quantum criticality. A similar suppression occurs in the resistivity measurements [Fig.~\ref{CRFiled}(b) and (c)], where $T_{\rm{N}}$ and $T_{\rm{M}}$ also move to lower temperatures with increasing field. Here, $T_{\rm{N}}$ also vanishes entirely above 6~T, suggesting the suppression of magnetic order, and the low temperature resistivity can be described by Fermi liquid behavior (dashed lines).

	\begin{figure}
		\includegraphics[angle=0,width=0.49\textwidth]{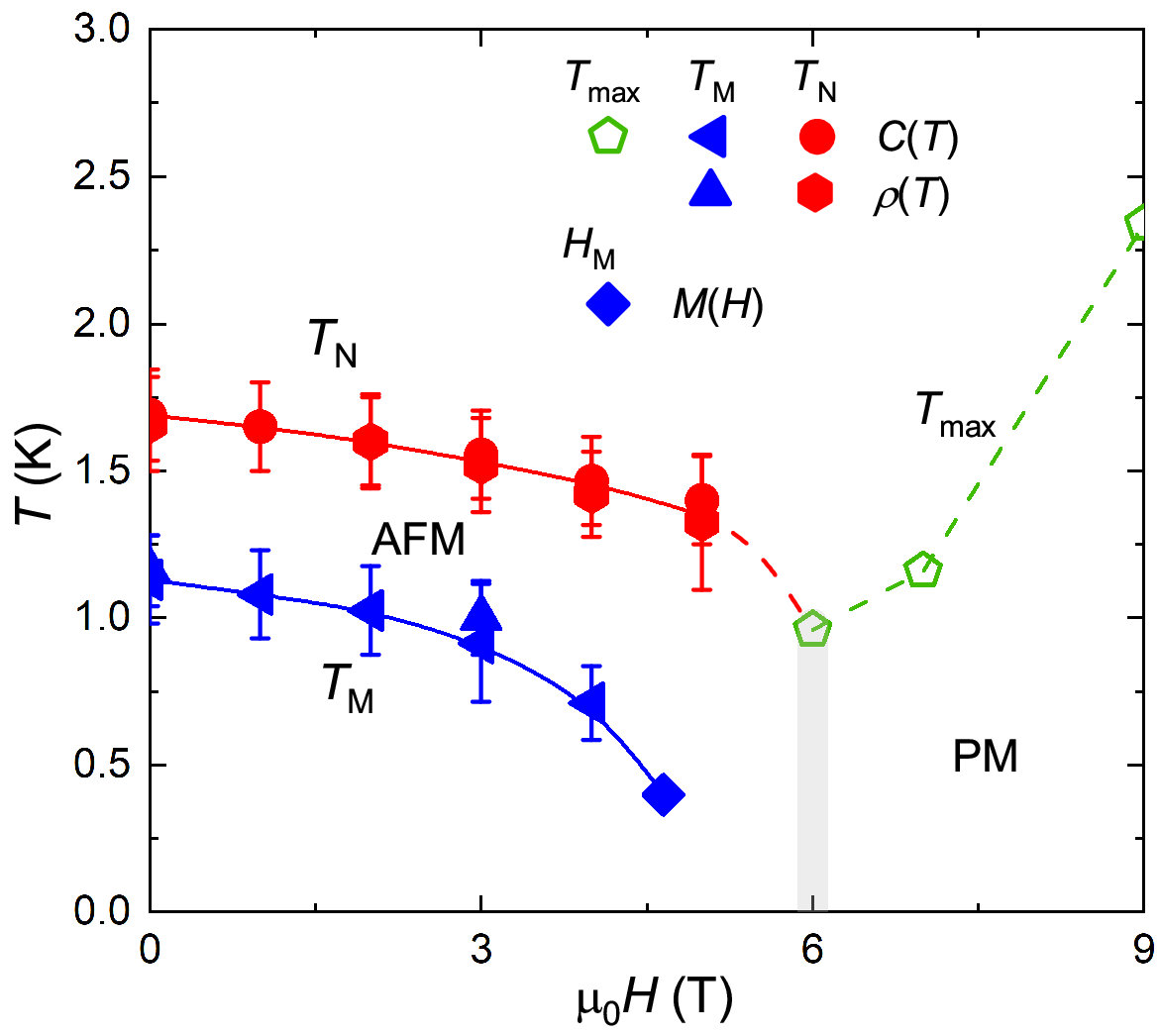}
		\vspace{-12pt} \caption{\label{HTPD} Magnetic field-temperature phase diagram of CePdIn for magnetic fields applied along the $c$-axis deduced from heat capacity, resistivity, and magnetization measurement. The vertical light gray line indicates the boundary between the antiferromagnetic state and the paramagnetic state. Error bars correspond to transition widths.}
		\vspace{-12pt}
	\end{figure}
	
	Based on the specific heat, resistivity, and magnetization measurements, we construct the magnetic field-temperature phase diagram of CePdIn, as shown in Fig.~\ref{HTPD}. When a magnetic field is applied along the $c$-axis, both transitions are suppressed; $T_{\rm{N}}$ shifts to lower temperatures until it vanishes abruptly above 6~T, and $T_{\rm{M}}$ follows a similar trend. At high-fields ($H > 6$~T), where no antiferromagnetic anomalies are observed, the system is likely in a spin-polarized state, where the resistivity shows Fermi liquid behavior.
	
	\subsection{B. Effect of hydrostatic pressure on the electrical resistivity and specific heat}
	
\begin{figure}
		\includegraphics[angle=0,width=0.49\textwidth]{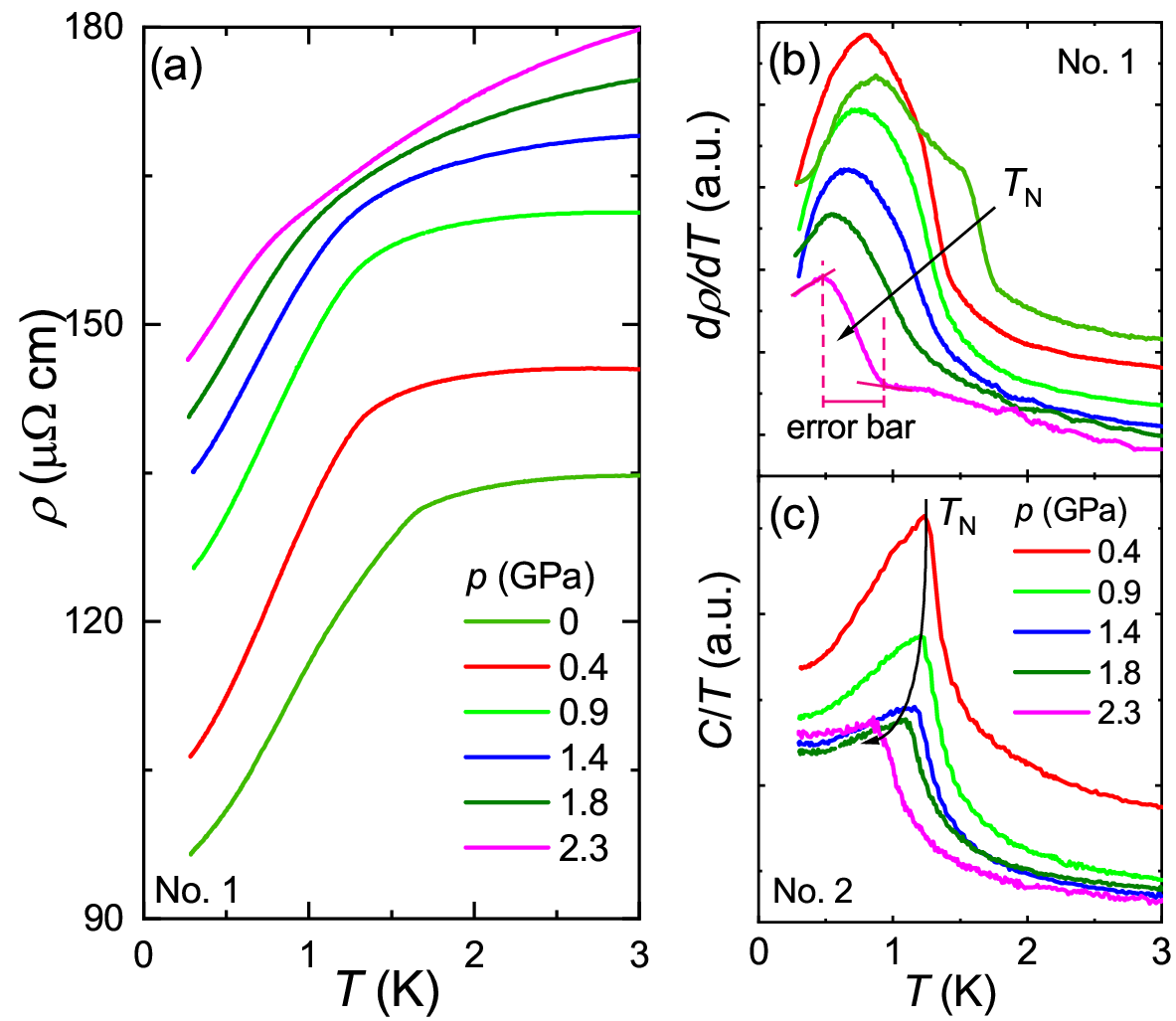}
		\vspace{-12pt} \caption{\label{RCT_p}  Temperature dependence of (a) $\rho(T)$ of sample No. 1, (b) d$\rho(T)$/d$T$, and (c) ac specific heat as $C/T$ of sample No. 2,  of CePdIn under various applied pressures up to 2.3 GPa, measured in a piston-cylinder pressure cell. The error bar in (b) at a selected pressure reflects the finite width of the transition.}
		\vspace{-12pt}
	\end{figure}

	To investigate the effects of pressure and to look for whether there is quantum criticality in CePdIn, we performed both resistivity and ac calorimetry measurements under hydrostatic pressure using a piston-cylinder cell, as shown in Fig.~\ref{RCT_p}. The $T_{\rm{N}}$ systematically shifts to lower temperatures with increasing pressure, as evidenced by both  $\rho(T)$ [Fig.~\ref{RCT_p}(a)] and its temperature derivative d$\rho(T)$/d$T$ [Fig.~\ref{RCT_p}(b)]. The ac calorimetry data [Fig.~\ref{RCT_p}(c)] on a different sample (No. 2) confirms the pressure evolution of $T_{\rm{N}}$. The lower transition $T_{\rm{M}}$ remains unresolved in both techniques, probably due to this being a weak feature, that becomes less clear with the pressure-induced broadening. Note that the maximum pressure of the piston-cylinder cell is insufficient to completely suppress $T_{\rm{N}}$. 
	
	\begin{figure}
		\includegraphics[angle=0,width=0.45\textwidth]{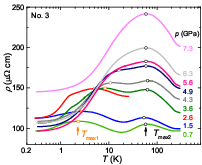}
		\vspace{-12pt} \caption{\label{RT_pDAC} Temperature dependence of $\rho(T)$ of CePdIn sample No. 3 measured under various pressures in a diamond-anvil cell. The arrows indicate the low- temperature and high- temperature maximums, $T_{\rm{max1}}$ and  $T_{\rm{max2}}$, whereas the open squares and circles show their respective positions at different pressures.}
		\vspace{-12pt}
	\end{figure}
	
	To probe the behavior of CePdIn at higher pressures,   $\rho(T)$ was measured  in a DAC, and the data at higher temperatures is shown in Fig.~\ref{RT_pDAC}. At low pressures, $\rho(T)$ exhibits two maxima ($T_{\rm{max1}}$ and $T_{\rm{max2}}$). The high-temperature maximum $T_{\rm{max2}}$ displays a weak pressure dependence, while the low-temperature one $T_{\rm{max1}}$ shifts upward with pressure and coalesces with $T_{\rm{max2}}$ near 6.3~GPa. The lower peak $T_{\rm{max1}}$ is likely associated with coherent Kondo scattering. Since pressure enhances the hybridization between the $f$-electrons and conduction electrons, the coherent Kondo state develops at higher temperatures under compression, in line with our results. In contrast, $T_{\rm{max2}}$ arises from inelastic scattering between conduction electrons and the excited crystal-electric-field (CEF) levels of the $f$-ions. As the CEF splittings are generally much less sensitive to pressure, $T_{\rm{max2}}$ exhibits only a weak pressure dependence. The merging of resistivity maxima signals a pressure-driven electronic crossover from a heavy fermion state with a small $T_{\rm{K}}$ to a degenerate state with an enhanced $T_{\rm{K}}$ that is of the same order as the CEF splitting. Similar behavior has been reported in other heavy fermion compounds, such as CeCu$_2$Si$_2$ \cite{PRBbb1984}, CeCu$_2$Ge$_2$ \cite{JMMMev1998}, and CeAu$_2$Si$_2$ \cite{PRXrz2014}.
	
\begin{figure}
		\includegraphics[angle=0,width=0.49\textwidth]{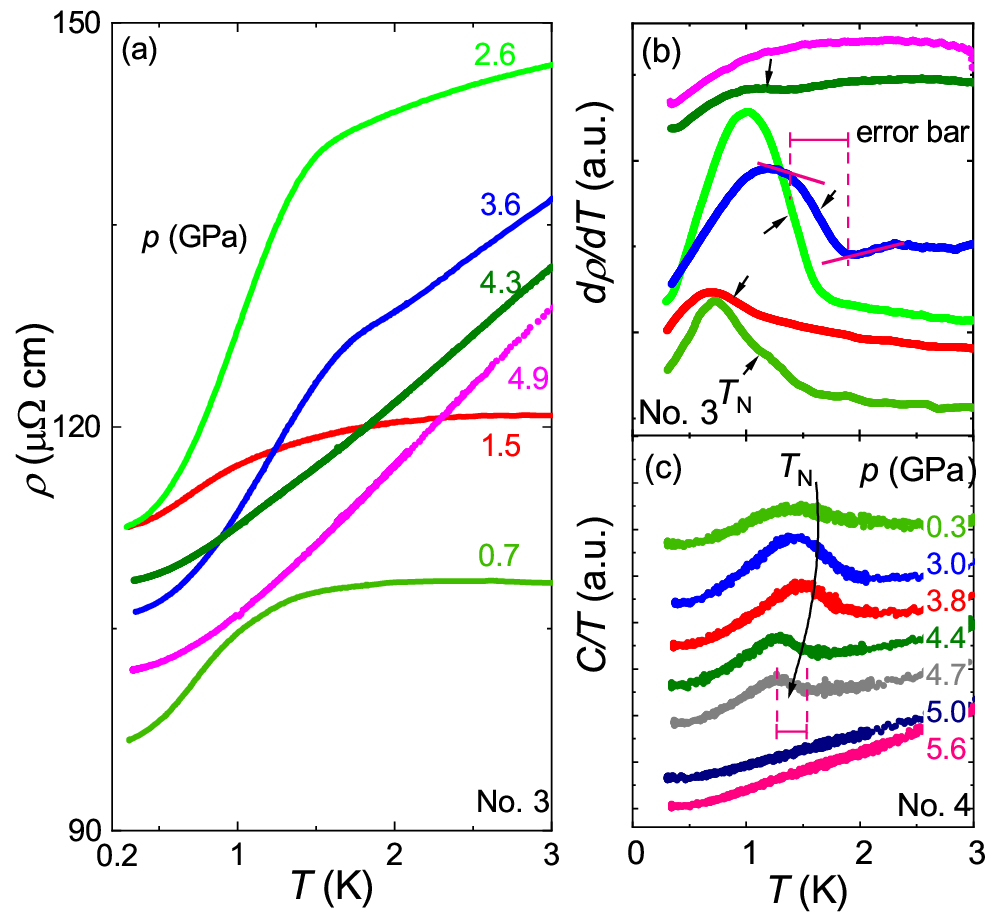}
		\vspace{-12pt} \caption{\label{lowTRT_DAC}  Low temperature  (a) $\rho(T)$ of sample No. 3,  the corresponding (b) d$\rho(T)$/d$T$, and ac specific heat as $C/T$ of sample No.4, of CePdIn measured under various pressures  in a DAC.  The arrows in (b) mark the  antiferromagnetic transition temperature $T_{\rm{N}}$, and arrow in (c) shows the evolution of $T_{\rm{N}}$ with pressure. The error bars in (b) and (c) at selected pressures reflect the finite width of the transition.}
		\vspace{-12pt}
	\end{figure}

	To track the pressure evolution of the magnetic ordering, we examine the low-temperature resistivity $\rho(T)$ and its derivative in Figs.~\ref{lowTRT_DAC}(a) and (b), respectively. The $T_{\rm{N}}$ exhibits a non-monotonic pressure dependence, whereby initially $T_{\rm{N}}$ decreases with pressure, reaching $\approx 0.9$~K at 1.5~GPa. Upon further compression, $T_{\rm{N}}$ increases and reaches a maximum at 3.6~GPa. Beyond 3.6~GPa, $T_{\rm{N}}$ decreases weakly and abruptly vanishes between 4.3 and 4.9~GPa. Ac calorimetry measurements $C(T)/T$ in the DAC [Fig.~\ref{lowTRT_DAC}(c)] show consistent behaviors, in that $T_{\rm{N}}$ reaches a maximum at 3.8~GPa, and suddenly disappears around 4.7--5.0~GPa. The resistivity reaches its maximum at 2.6~GPa, likely due to enhanced spin fluctuations associated with the suppression of magnetic order below this pressure. Note that the aforementioned merging of $T_{\rm{max1}}$ and $T_{\rm{max2}}$ occur at higher pressure than that at which the magnetic order disappears.
	
	\begin{figure}
		\includegraphics[angle=0,width=0.45\textwidth]{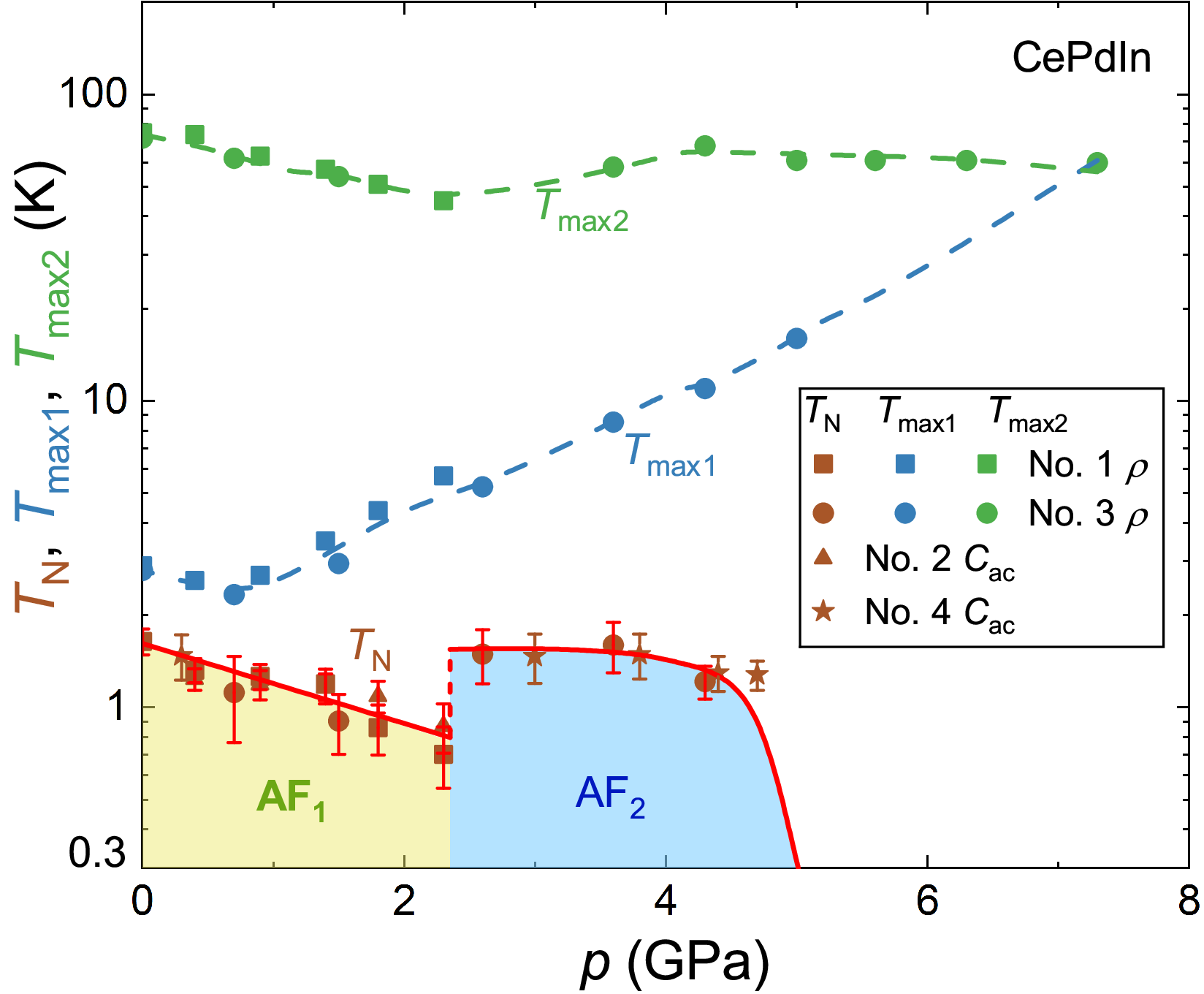}
		\vspace{-12pt} \caption{\label{pTPD} Pressure - temperature phase diagram of CePdIn, where the temperature is on a logarithmic scale. $T_{\rm{max1}}$ and $T_{\rm{max2}}$ are the positions of the lower and higher temperature resistivity maxima, respectively. $T_{\rm{N}}$ was derived from resistivity and ac calorimetry measurements. Error bars correspond to transition widths.}
		\vspace{-12pt}
	\end{figure}
	
	Combining resistivity and ac calorimetry measurements, we construct the pressure-temperature phase diagram of CePdIn, which is shown on a logarithmic temperature scale in Fig.~\ref{pTPD}. Here, $T_{\rm{max2}}$ shows a slight variation with  pressure, while $T_{\rm{max1}}$ initially decreases slightly below 1~GPa, then increases steadily, merging with $T_{\rm{max2}}$ above 6.3~GPa. Below 2.6~GPa (AF$_1$ phase), $T_{\rm{N}}$ is suppressed with increasing pressure At 2.6~GPa, $T_{\rm{N}}$ shows an abrupt enhancement, signaling a transition to a distinct magnetic phase labelled AF$_2$. In the AF$_2$ regime, $T_{\rm{N}}$ varies weakly with pressure before disappearing abruptly near 5~GPa. Note that the pressure in the DAC was calibrated at room temperature. Upon cooling, a slight increase in pressure is generally expected in a DAC. Since in situ pressure determination at low temperature is not available in the present setup, this effect is not explicitly included in the pressure–temperature phase diagram. Nevertheless, a direct comparison between measurements performed in a piston–cylinder cell and those carried out in a DAC reveals no visible deviation in the evolution of $T_{\rm{N}}$ with pressure (Fig.~\ref{pTPD}), indicating that the pressure change upon cooling in the DAC is small and does not qualitatively affect the phase diagram.
	
	\subsection{C. Distinct magnetic phases AF$_1$ and AF$_2$}
	
	\begin{figure}
		\includegraphics[angle=0,width=0.45\textwidth]{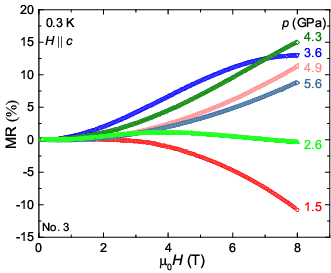}
		\vspace{-12pt} \caption{\label{MR} Magnetoresistivity of CePdIn sample No. 3 under different pressures, measured under different pressures at 0.3 K with the magnetic field applied along the $c$-axis.}
		\vspace{-12pt}
	\end{figure}
	
	To probe the distinct nature of the two magnetic phases, we measured the magnetoresistivity (MR) of CePdIn under applied magnetic fields at various pressures, which is displayed in Fig.~\ref{MR} at $T = 0.3$~K across different pressure regimes.  At 1.5~GPa in the AF$_1$ phase, the MR is overall negative, except for a small positive value below 2~T. With increasing pressure, this negative MR is gradually suppressed and becomes positive. At 2.6~GPa, which is at the crossover from AF$_1$ to AF$_2$, the MR shows visible non-monotonic field dependence and displays a local maximum at 3.5~T. In the AF$_2$ phase ($>2.6$~GPa), CePdIn shows a positive MR for the whole field range.  Once the magnetic order is suppressed by pressure at 4.9~GPa and 5.6~GPa,  the MR follows a conventional quadratic field dependence, consistent with normal metallic behavior. Such an evolution of the MR provides supporting evidence for a change of the underlying electronic states between the AF$_1$ and AF$_2$ phases.

	\begin{figure}
		\includegraphics[angle=0,width=0.49\textwidth]{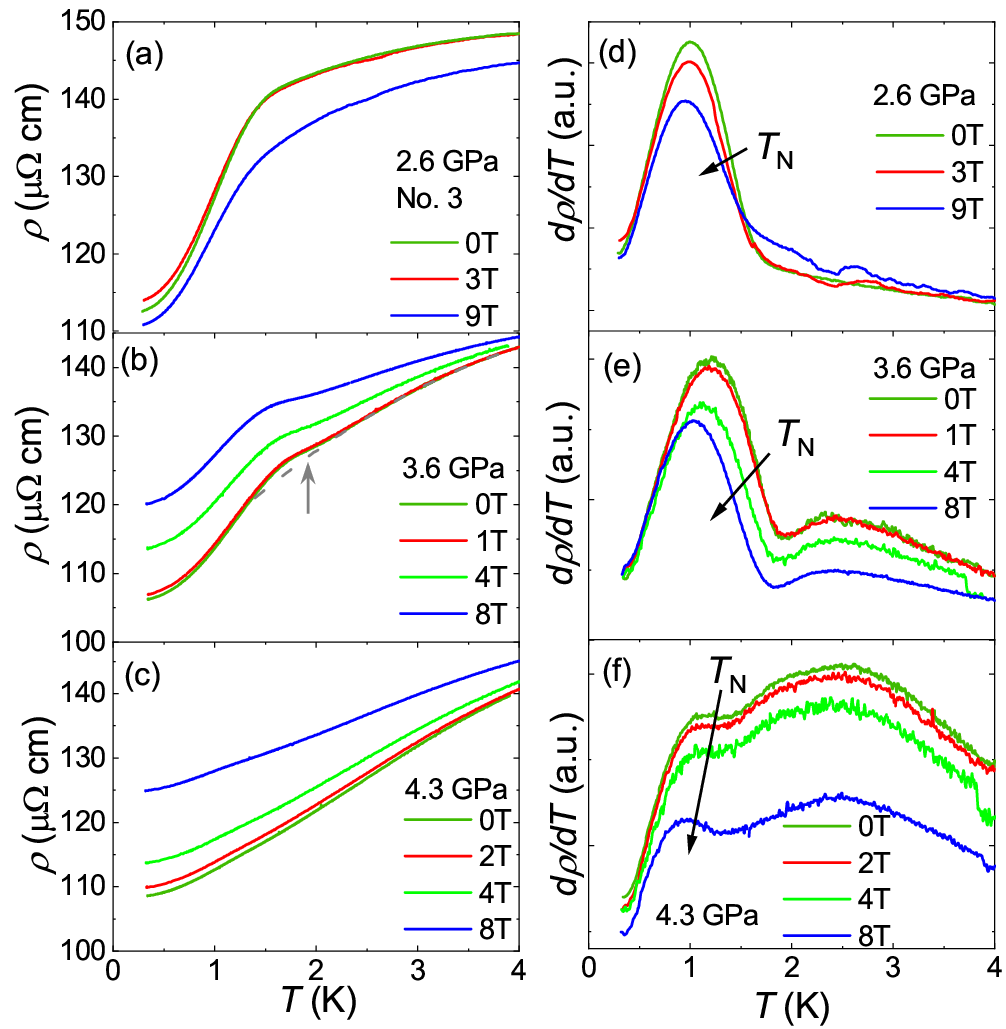}
		\vspace{-12pt} \caption{\label{RTfid} Temperature dependence of $\rho(T)$ of CePdIn sample No. 3 in various fields at (a) 2.6~GPa, (b) 3.6~GPa and (c) 4.3~GPa. The corresponding derivatives d$\rho(T)$/d$T$ are shown at (d) 2.6~GPa, (e) 3.6~GPa, and (f) 4.3~GPa. The dashed line in (b) represents a simple polynomial fit to the high-temperature regime above the magnetic ordering temperature. The gray arrow highlights the slight enhancement in resistivity as the system enters the magnetic phase.}
		\vspace{-12pt}
	\end{figure}
	
	Figure~\ref{RTfid} displays  $\rho(T)$ and its derivative d$\rho(T)$/d$T$ in the AF$_2$ phase. In a 8~T magnetic field $T_{\rm{N}}$ is slightly suppressed but the transition can still be resolved. This contrasts sharply with the field dependence in the  AF$_1$ phase, where the magnetic order is less robust despite the comparable values of $T_{\rm{N}}$, being suppressed in a 6~T field. These suggest that the AF$_1$ and AF$_2$ phases correspond to distinct antiferromagnetic orders, accompanying the evolution of the electronic state.
	
	\section{VI. DISCUSSION}
	
	CePdIn exhibits a more three-dimensional crystal structure compared to its isostructural counterpart \mbox{CePdAl}, which may result in significantly weaker geometric frustration. Magnetic field studies reveal a gradual suppression of the antiferromagnetic phase in CePdIn, contrasting sharply with the complex temperature-field phase diagram of CePdAl that hosts multiple metamagnetic transitions, together with a possible spin-liquid phase \cite{PRLsl2017,PRBzh2016}. Notably, such field-induced anomalies are also commonly observed in other frustrated ZrNiAl-type systems such as YbAgGe \cite{PRLty2013}, YbPtIn \cite{PRLme2006}, and CePtPb \cite{CPP2019}, but are  absent in CePdIn. Due to the large number of nearly-degenerate ground states in frustrated systems, they often host multiple field-driven phases, both in the aforementioned intermetallics, as well as insulating quantum antiferromagnets \cite{JPCMah2004}, and therefore the lack of such phenomena in CePdIn is consistent with weaker frustration. Near the critical field where AFM order vanishes, the specific heat $C/T$ does not show a zero-temperature divergence, but rather a hump, a response that parallels observations in CePdAl \cite{PRLsl2017}, YbPtIn \cite{PRLme2006}, and CePtPb \cite{CPP2019}. This hump reflects the Zeeman splitting of the CEF ground-state doublet. Above the critical field, the Zeeman splitting exceeds the magnetic coupling, driving a transition from the antiferromagnetic state to the paramagnetic state and resulting in a first-order phase transition. As the magnetic field enhances the CEF ground-state splitting, the characteristic maximum in the heat capacity shifts to higher temperatures with increasing field.
	
	Under pressure, CePdIn exhibits a non-monotonic evolution of $T_{\rm{N}}$ followed by its abrupt disappearance (i.e., a first order phase transition), and therefore does not follow the straightforward Doniach picture for the competition between the Kondo and RKKY interactions \cite{DONIACH1977231}. Similar sudden collapses of magnetic order have been reported in CeRu$_2$Al$_{10}$ \cite{JPSJ.78.123705} and CeOs$_2$Al$_{10}$ \cite{JPSJ.80.064709}. The latter shows particular parallels to CePdIn, namely in CeOs$_2$Al$_{10}$, $T_{\rm{N}}$ decreases only modestly from 28~K at ambient pressure to 25~K near the critical pressure $p_{\rm{c}}$, before vanishing abruptly. Intriguingly, neutron diffraction studies suggest that while $T_{\rm{N}}$ remains nearly unchanged in CeOs$_2$Al$_{10}$, the volume fraction of the AFM phase gradually diminishes to zero at $p_{\rm{c}}$ \cite{JPSJ.80.064709}. However, further experimental and theoretical investigations to unravel the microscopic mechanisms governing such discontinuous quantum phase transitions are required.
	
	The AF$_1$ and AF$_2$ phases exhibit strikingly different responses to magnetic fields despite their similar values of  $T_{\rm{N}}$. While $T_{\rm{N}}$ in AF$_1$ is strongly suppressed by applied fields, it remains significantly more robust in AF$_2$, highlighting that these phases separated near 2.6~GPa correspond to distinct magnetically ordered states. This is further evidenced by transport behavior, where in the AF$_2$ phase, $\rho(T)$ is slightly enhanced below $T_{\rm{N}}$ [Fig.~\ref{RTfid} (b)]. One possibility is that this corresponds to the opening of a gap associated with spin-density wave type magnetism, which might suggest that the AF$_2$ phase is of the itinerant-type. In this case, the reduced frustration could lead to the delocalization of the $4f$ electrons within the magnetically ordered phase \cite{PBqs2006}, in contrast to CePdAl where it is proposed that frustration leads to a Kondo destruction type quantum criticality under pressure in the paramagnetic state \cite{NP2019CPA}. On the other hand, for CePdIn at ambient pressure, no magnetic Bragg peaks were  observed with neutron diffraction \cite{Klicpera2017}, indicating a significantly reduced ordered moment, which suggests that even in the AF$_1$ phase the Kondo effect significantly quenches the Ce-$4f$ moments. Alternatively, strong critical antiferromagnetic fluctuations or a suppression of Kondo coherence  could also contribute to an increase in resistivity around $T_{\rm{N}}$ in the AF$_2$. Additional sensitive microscopic probes of the magnetic order such as polarized neutron diffraction or muon-spin relaxation could be important for understanding the nature of the magnetic ordering in CePdIn. Chemical substitution can generate positive chemical pressure, functioning in a manner analogous to hydrostatic pressure. Consistent with this, substituting Pd with Ni reduces the unit-cell volume and suppresses the antiferromagnetic order~\cite{JPCMmk2013}. Notably, however, no counterpart to the AF$_2$ phase identified in our work has been observed, suggesting that chemical pressure and hydrostatic pressure influence the system in different ways.

	\section{V. CONCLUSION}
	
	In summary, we have established the magnetic field–temperature and pressure–temperature phase diagrams of CePdIn through  calorimetric and electrical transport measurements. At ambient pressure, both the antiferromagnetic transition $T_{\rm{N}}$ and  additional low temperature transition $T_{\rm{M}}$ are suppressed by magnetic fields, and abruptly disapppear above 6~T, where the system likely enters the spin polarized state. Under hydrostatic pressure, $T_{\rm{N}}$ exhibits a non-monotonic evolution; it initially decreases before undergoing an abrupt enhancement near 2.6~GPa, followed by a weak change and eventual sudden disapperance at ~5~GPa. Notably, the high-pressure AF$_2$ phase (above 2.6~GPa) is more robust in $c$-axis magnetic fields compared to the low-pressure AF$_1$ phase. This suggests that these are two distinct magnetically ordered phases, and this change may be driven by changes of the underlying electronic state, due to enhanced Kondo hybridization with pressure. The effects of pressure and field tuning on CePdIn  are markedly different from CePdAl, highlighting that both reduced magnetic anisotropy and weaker geometric frustration, possibly driven by a more three dimensional structure, lead to greatly different quantum critical behaviors within a single family of materials. Such comparisons can also provide important information as to how frustration can be tuned in metallic Kondo lattice compounds.

\section{ACKNOWLEDGMENTS}

This work was supported by the National Key R\&D Program of China (Grant No. 2022YFA1402200 and No. 2023YFA1406303),  the National Natural Science Foundation of China (Grants No. 12174332, No. 12034017, No. W2511006, No. U23A20580, No. 12350710785, and No. 12204159), and the Zhejiang Provincial Natural Science Foundation of China (Grant No. LRG26A040001).

\end{document}